\documentclass[9pt,twocolumn,twoside]{optica}
\setboolean{shortarticle}{false}

\title{Dispersion-enabled quantum state control in integrated photonics}

\author[1]{Ryan P. Marchildon}
\author[1,2]{Amr S. Helmy}

\affil[1]{\small The Edward S. Rogers Department of Electrical and Computer Engineering,
University of Toronto, 10 King\textquoteright{}s College Road, Toronto,
Ontario M5S 3G4, Canada.}
\affil[2]{\small Institute for Optical Sciences, University of Toronto, 60 St. George Street, Toronto, Ontario M5S 3G4, Canada.}

\dates{Compiled \today}

\ociscodes{(270.5585) Quantum information and processing; (130.0130) Integrated optics; (230.3990) Micro-optical devices.}

\doi{\url{http://dx.doi.org/10.1364/optica.XX.XXXXXX}}

\begin{abstract}
Integrated optics has brought unprecedented levels of stability and performance to quantum photonic circuits. However, integrated devices are not merely micron-scale equivalents of their bulk-optics counterparts. By exploiting the inherently dispersive characteristics of the integrated setting, such devices can play a remarkably more versatile role in quantum circuit architectures. We show this by examining the implications of linear dispersion in an ordinary directional coupler. Dispersion unlocks several novel capabilities for this device, including in-situ control over photon spectral and polarization entanglement, tunable photon time-ordering, and entanglement-sensitive two-photon coincidence generation. Also revealed is an ability to maintain perfect two-photon anti-coalescence while tuning the interference visibility, which has no equivalent in bulk-optics. The outcome of this work adds to a suite of state engineering and characterization tools that benefit from the advantages of integration. It also paves the way for re-evaluating the possibilities offered by dispersion in other on-chip devices.
\end{abstract}

\setboolean{displaycopyright}{true}

\begin{document}

\maketitle
\thispagestyle{fancy}
\ifthenelse{\boolean{shortarticle}}{\abscontent}{}

\section{Introduction}

\par {The quantum properties of light can unlock a variety of enhanced and novel technological capabilities. Among these are secure communications \cite{Gisin_RevModPhys_2002, Lo_PRL_2012}, non-classical simulation \cite{Ma_NaturePhysics_2011}, non-local imaging \cite{Lemos_Nature_2014}, and pathway-selective exciton spectroscopy \cite{Schlawin_NatureComms_2013}. Such quantum photonic technologies have traditionally been implemented on the bench-top with discrete optical components. More recently, the need for improved scalability has fuelled widespread interest in the development of on-chip quantum circuits. Much of this work has concentrated on the generation \cite{Luxmoore_SciRep_2013, Davanco_APL_2012, Matsuda_SciRep_2012, Horn_SciRep_2013}, manipulation \cite{Matthews_NaturePhotonics_2009, Sansoni_PRL_2010, Shadbolt_NaturePhotonics_2011, Tanzilli_LaserPhotRev_2011, Wang_OptComms_2014, Silverstone_NaturePhotonics_2014}, and detection \cite{Pernice_NatureComms_2012, Reithmaier_SciRep_2013} of entangled \cite{Horodecki_RevModPhys_2009} photon pairs, often with the goal of replicating tasks previously performed using bulk-optics. However, integrated optical components can exhibit highly wavelength-dependent (i.e. dispersive) behaviour compared to their bulk-optics counterparts, and investigating whether this leads to functionalities not previously available is also an important objective. Such dispersion has been shown to provide unprecedented tailorability over the properties of two-photon states generated by engineered nonlinear interactions  \cite{Abolghasem_OL_2009, Eckstein_PRL_2011, Kang_JOSAB_2014} in integrated waveguides. Here we consider new ways of leveraging dispersion for the manipulation of two-photon states and their correlation properties. 

\par Directional couplers are a common building block of integrated quantum circuits whose dispersion properties have yet to be fully exploited. They are typically implemented through the evanescent coupling of two identical waveguides and are characterized by a power splitting ratio $\eta(\lambda)$. Their primary role has been to serve as on-chip beamspitters, often to mediate quantum interference \cite{Matthews_NaturePhotonics_2009, Sansoni_PRL_2010, Shadbolt_NaturePhotonics_2011, Wang_OptComms_2014}.  Due to the presence of dispersion in $\eta(\lambda)$, these same couplers can also act as a wavelength demultiplexer (WD) for specific sets of non-degenerate wavelengths, without relying on waveguide modal mismatch. In fact, dispersion can cause the coupler's behaviour to transition between `ideal' beamsplitter operation and `ideal' WD operation in response to either the properties of the quantum state or systematic shifts to the coupling strength. The implications this has for two-photon state manipulation has yet to be studied. We show that this attribute of directional couplers grants them a versatile set of new functionalities, which includes the post-selective tuning of spectral entanglement, entanglement-sensitive coincidence detection, and the ability to maintain perfect anti-coalescence while allowing full tunability over the two-photon interference visibility.

\par In what follows we use symmetric 2x2 directional couplers as an example of quantum state engineering in integrated photonic systems without loss of generality. As such, an essential step is to parameterize the coupler's response for the two-photon state in terms of generic dimensionless variables that can be mapped to any combination of coupler and state properties. The details of this parametrization are described in the Methods section, but we introduce the key definitions here. Suppose two single-mode waveguides are coupled over a length $L$, such as in Figure~\ref{Fig:CouplerResponse}(a). For symmetric rectangular waveguides, this leads to a power splitting ratio of  $\eta(\lambda) = \cos^{2}\left(\kappa(\lambda)L\right)$, where $\kappa(\lambda)$ is the coupling strength \cite{Yariv_JQE_1973, Taylor_ProcIEEE_1974}. The wavelength dependence of the coupler is usually sufficiently described by its first-order coupler dispersion $\textrm{M} = \mathrm{d}\kappa(\lambda)L/\mathrm{d}\lambda$ at a reference wavelength $\lambda_{00}$, together with the value of $\eta(\lambda_{00})$. Let $\lambda_{01}$ and $\lambda_{02}$ be the central wavelengths of a photon pair that evolves through this coupler, with $\Lambda = \vert \lambda_{02} - \lambda_{01} \vert$ giving the non-degeneracy. Defining $\Delta\eta = \big\vert \eta(\lambda_{02}) - \eta(\lambda_{01}) \big\vert$ allows the coupler response to be classified as beamsplitter-like for $\Delta\eta \rightarrow 0$ or WD-like for $\Delta\eta \rightarrow 1$. The space of all possible $\Delta\eta$ is spanned by $\eta(\lambda_{00})$ and the dimensionless product $\textrm{M}\Lambda$. This has been plotted in Figure~\ref{Fig:CouplerResponse}(b), assuming $\lambda_{00} =  \left( \lambda_{01} +  \lambda_{02} \right)/2$ and negligible higher-order coupler dispersion (see Methods). This plot provides a useful guide for relating the results of this paper to transitions between beamsplitter and WD behaviour. A special condition, $\eta(\lambda_{01}) + \eta(\lambda_{02}) = 1$, occurs along the lines $\eta(\lambda_{00})=0$ and $\mathrm{M}\Lambda = \pi/2$, and corresponds to the splitting ratios $\eta(\lambda_{01})$ and $\eta(\lambda_{02})$ being anti-symmetric about the 50:50 splitting value $\eta = 0.5$. This will turn out to have important implications for tasks involving photon anti-coalescence.

\begin{figure}[t!]
\centering
\includegraphics[width=0.95\linewidth]{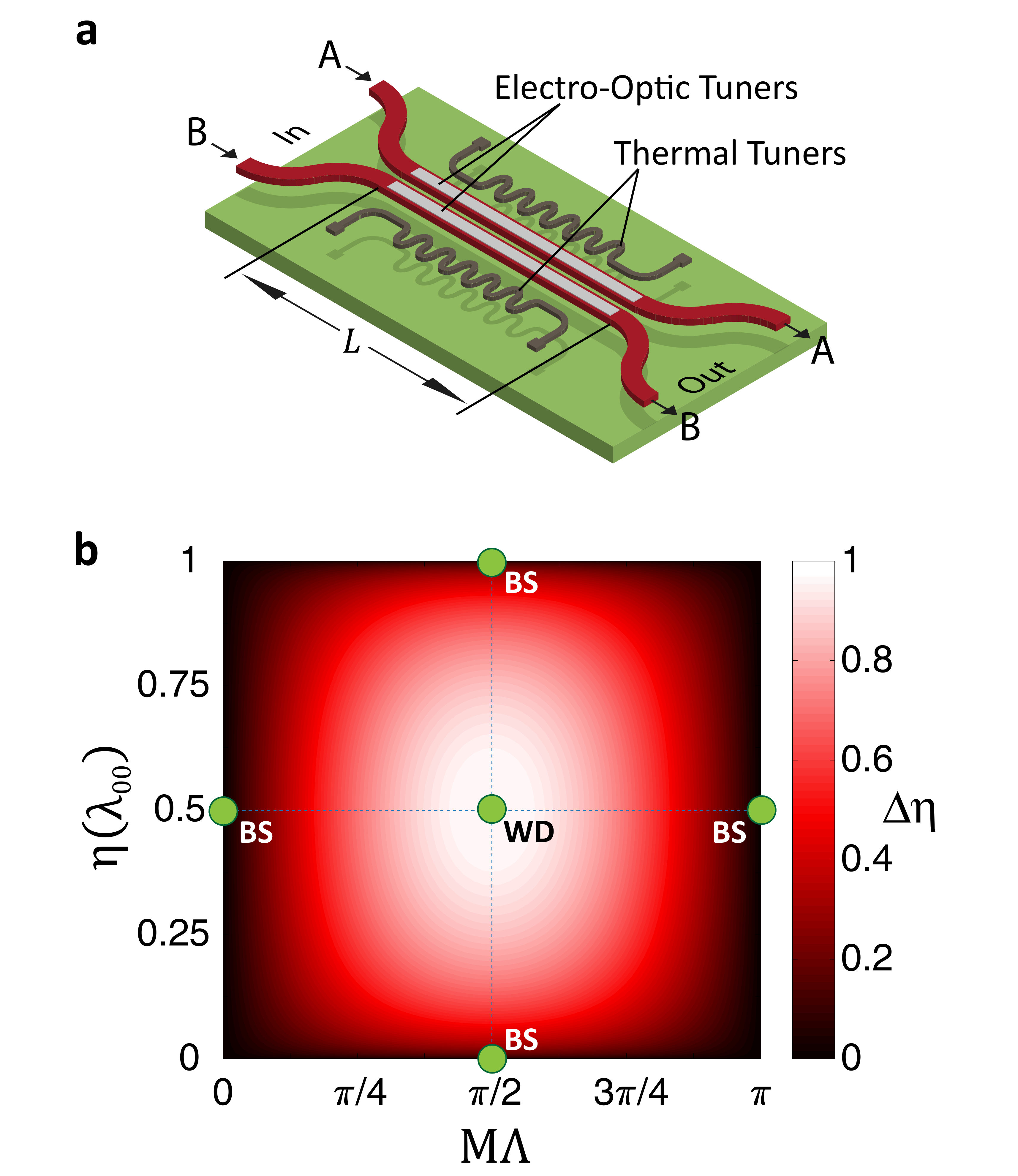}
\caption{\small \textbf{Navigating the coupler response.} (\textbf{a}) Depiction of a generic two-port directional coupler, shown with simple implementations of thermal and electro-optic tuning for in-situ control over $\eta(\lambda_{00})$. (\textbf{b}) Map of possible coupler responses to a two-photon input state, as characterized by $\Delta\eta$. The coordinates labelled BS denote 50:50 beamsplitter behaviour, while WD denotes perfect demultiplexing of central wavelengths $\lambda_{01}$ and $\lambda_{02}$.}
\label{Fig:CouplerResponse}
\end{figure}

\section{Dispersion-Enabled Capabilities} 

\subsection{Tunable Spectral Entanglement}
\par Suppose two non-degenerate photons enter a directional coupler from a single input port, so that the input state takes the form $\vert \psi \rangle_{\textrm{in}} = \left\vert\lambda_{01} \right\rangle_{j}\left\vert\lambda_{02} \right\rangle_{j}$ where $j \in \{A,B\}$. The two-photon state at the output of the coupler is then post-selected for outcomes where the photons exit from different waveguides (i.e. separated). Depending on the coupler response, the output waveguide taken by a given photon can reveal information about that photon's spectral properties, which in turn alters the spectral entanglement of the post-selected output state. A WD-like response with $\Delta\eta = 1$ pre-determines which photon emerges from each output port. This leads to an output state of the form $\vert \psi \rangle_{\textrm{out}} = \vert \lambda_{01}\rangle_{A} \vert \lambda_{02} \rangle_{B}$ (or $\vert \psi \rangle_{\textrm{out}} = \vert \lambda_{01}\rangle_{B} \vert \lambda_{02} \rangle_{A}$, depending on the input port), where entanglement of the central wavelengths is lost. On the other hand, a beamsplitter-like response with $\Delta\eta = 0$ leads to the superposition $\vert \psi \rangle_{\textrm{out}} = \left[ \vert \lambda_{01}\rangle_{A} \vert \lambda_{02} \rangle_{B} + \vert \lambda_{01}\rangle_{B} \vert \lambda_{02} \rangle_{A}\right]/\sqrt{2}$, where the full spectral entanglement of the input state is retained. By controlling $\Delta\eta$ through the selection of $\textrm{M}$ or $\eta(\lambda_{00})$ (and thus controlling, effectively, the amount of spectral information known about the output state), a directional coupler can select any level of entanglement between these extremes.

\begin{figure}[b!]
\centering
\includegraphics[width=\linewidth]{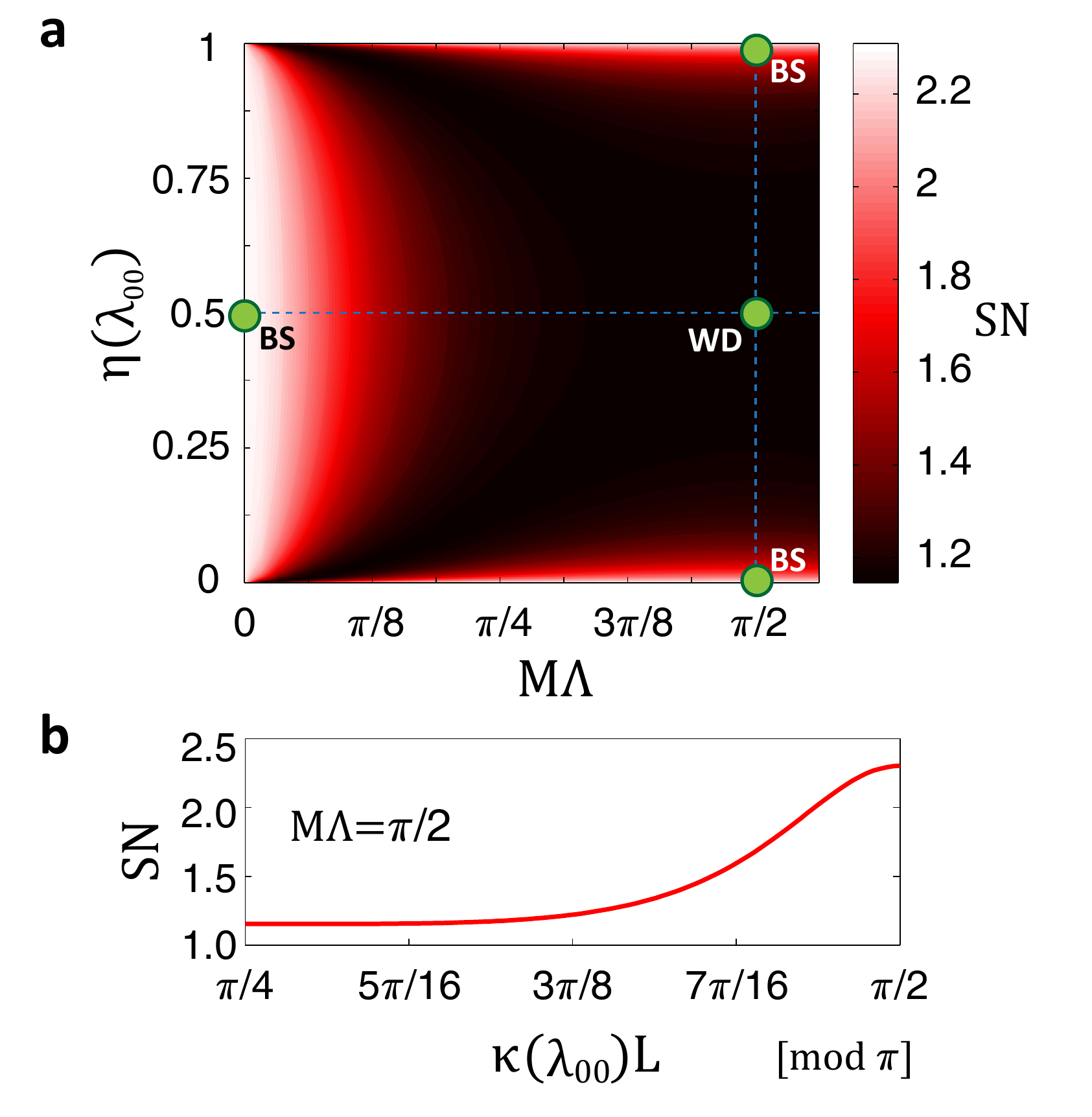}
\caption{\small \textbf{Tunability of ouput state entanglement.}  (\textbf{a})~Dependence of Schmidt Number on the coupler response for post-selected outcomes where the photons are found in different waveguides. The maximum value of  $\textrm{SN}=2.31$ corresponds to the input state entanglement. (\textbf{b})~Slice along $\mathrm{M}\Lambda = \pi/2$, plotted in terms of $\kappa(\lambda_{00})$.}
\label{Fig:TunableEntanglement}
\end{figure}

\par Figure~\ref{Fig:TunableEntanglement} shows how the choice of $\textrm{M}$ and $\eta(\lambda_{00})$ can tailor spectral entanglement in the post-selected output state (see Methods for calculation details). Spectral entanglement has been quantified using the Schmidt Number (SN) \cite{Parker_PRA_2000, Humble_PRA_2008}, which has a minimum value of unity in the absence of entanglement, and increases with greater entanglement. The input state used in this example has $\Lambda = \textrm{10~nm}$, $\textrm{SN}=2.31$, and equal FWHM intensity bandwidths of $\Delta\lambda = \textrm{1~nm}$ for the photon marginal spectra. It is modelled after a Type-I spontaneous parametric downconversion (SPDC) process \cite{Yang_PRA_2008} with a degeneracy wavelength of $\lambda_{00} = \textrm{1550~nm}$ and pump bandwidth of $\Delta\lambda_{\textrm{P}}=\textrm{0.25~nm}$. As the coupler response moves away from the beamsplitter-like coordinates and towards the WD-like coordinate at ($\mathrm{M}\Lambda=\pi/2$, $\eta(\lambda_{00})= 0$), the Schmidt Number of the output state smoothly transitions from its input value of $\textrm{SN}=2.31$ down to a value of $\textrm{SN} \simeq 1.15$. Note that some spectral entanglement remains at the WD-like coordinate even though the output paths reveal the central wavelengths. This is because the photon spectra are still inherently anti-correlated about their central wavelengths, due to energy and momentum conservation in the pair generation process. Such residual entanglement vanishes as $\Delta\lambda \rightarrow 0$.

\par In-situ tuning of the Schmidt Number becomes possible through active control of $\eta(\lambda_{00})$. Effectively, this prepares states of the form $\vert \psi \rangle_{\textrm{out}} = \left[ \vert \lambda_{01}\rangle_{A} \vert \lambda_{02} \rangle_{B} + \mu\vert \lambda_{01}\rangle_{B} \vert \lambda_{02} \rangle_{A} \right]/\sqrt{1 + \mu^2}$ with a tunable value of $\mu$. We emphasize that this tuning occurs post-generation, without requiring changes to pump bandwidth, nonlinear interaction length, or any other parameters affecting the photon pair generation process. This makes it particularly well-suited for tailoring spectral entanglement in a monolithically-integrated setting, in applications where the photons remain path-distinguishable. Control of $\eta(\lambda_{00})$, and thereby the Schmidt Number, can be achieved electro-optically or thermally by modifying the waveguide core-cladding index contrast to systematically shift $\kappa(\lambda_{00})$. Other potential tuning methods include the quantum-confined stark effect \cite{Wood_JLT_1988, Stohr_OQE_1993} and, for certain fiber-based coupler assemblies, a micrometer-controlled waveguide separation \cite{Digonnet_JQE_1982}. Operation along the line $\mathrm{M}\Lambda = \pi/2$ offers the most precise control over entanglement at any non-zero $\Lambda$. The value of $\mathrm{M}$ is fixed but can be tailored through judicious design of the coupler dimensions and material system. Note that since $\mathrm{M}$ scales with $L$, dispersion can be enhanced by increasing the 50:50 coupling length beyond its minimum necessary value of $L = \pi/\big(4 \kappa(\lambda_{00}) \big)$.

\par This tuning approach also provides control over polarization entanglement, since correlations in the spectral and polarization degrees of freedom are coupled \cite{Humble_PRA_2008}, except in the special case of maximal polarization entanglement. A state's polarization entanglement can be quantified using its concurrence $C$ \cite{Hill_PRL_1997, Wootters_PRL_1998}, with $C=0$ and $C=1$ indicating minimal and maximal entanglement respectively. As the state Schmidt Number increases, polarization entanglement tends to decrease, and vice-versa \cite{Humble_PRA_2008}. This inverse relation between SN and $C$ allows for the on-chip preparation of non-maximally entangled states ${\vert \psi \rangle = \left( \vert H, V \rangle + r\exp{i\phi} \vert V, H \rangle \right)/\sqrt{1+r^2}}$ with a tunable value of $r < 1$, with $r$ related to the concurrence by $C = 2r/\left(1+r^2\right)$. Such states offer significant advantages over maximally entangled states in certain applications such as closing the detection loophole in quantum nonlocality tests \cite{Christensen_PRL_2013}.

\par The tunable spectral entanglement we present may also have useful capabilities for two-photon spectroscopy \cite{Schlawin_JChemPhys_2013} and light-induced matter correlations \cite{Muthukrishnan_PRL_2004, Schlawin_PRA_2014}. In these applications, the time-ordering of when each photon reaches the sample can affect the two-photon absorption probability. This is because a particular two-photon transition can have pairings of absorption pathways corresponding to whether $\lambda_{01}$ or $\lambda_{02}$ is absorbed first. For some systems, when both time orderings are permitted by the incident light, these pathways destructively interfere to suppress the two-photon absorption probability, as is the case for two uncoupled two-level atoms \cite{Muthukrishnan_PRL_2004}. Such transitions can thus be selectively controlled by changing which time-orderings (and hence absorption pathways) are allowed. 

\par As illustrated in Figure~\ref{Fig:TimeOrdering}, control over the allowed time-orderings is achievable by placing a time delay in one path (e.g. path $A$) and tuning $\mu$ by tuning the coupler parameter $\eta(\lambda_{00})$. Suppose $\mu=0$ (Fig.~\ref{Fig:TimeOrdering}(b)) so that the post-selected state at the coupler output is $\vert \psi \rangle_{\textrm{out}} = \vert \lambda_{01}\rangle_{A} \vert \lambda_{02} \rangle_{B}$. In this case, $\lambda_{01}$ is always delayed relative to $\lambda_{02}$, hence only one set of time-ordered pathways is allowed. On the other hand, when $\mu=1$ (Fig.~\ref{Fig:TimeOrdering}(c)) so that $\vert \psi \rangle_{\textrm{out}} = \left[ \vert \lambda_{01}\rangle_{A} \vert \lambda_{02} \rangle_{B} + \vert \lambda_{01}\rangle_{B} \vert \lambda_{02} \rangle_{A}\right]/\sqrt{2}$, the delay is applied in superposition to either $\lambda_{01}$ or $\lambda_{02}$, and hence both sets of time-ordered pathways are allowed.

\par Such control over the time-ordering adds to the versatility of a single on-chip light source for manipulating and probing two-photon processes, such as controlling the degree to which bi-exciton transitions may be blocked \cite{Schlawin_NatureComms_2013}. Note that the ability to selectively excite a single absorption path (e.g. using $\mu=0$) is only possible with quantum light sources. Classical sources have no intrinsic time-ordering and hence will excite both paths equally (as with $\mu = 1$). A tunable dispersive coupler thus allows the sample’s behaviour for both the classical and non-classical conditions to be directly compared, without the need to change the light source and with virtually no disruption to the experimental setup.

\begin{figure}[t!]
\centering
\includegraphics[width=\linewidth]{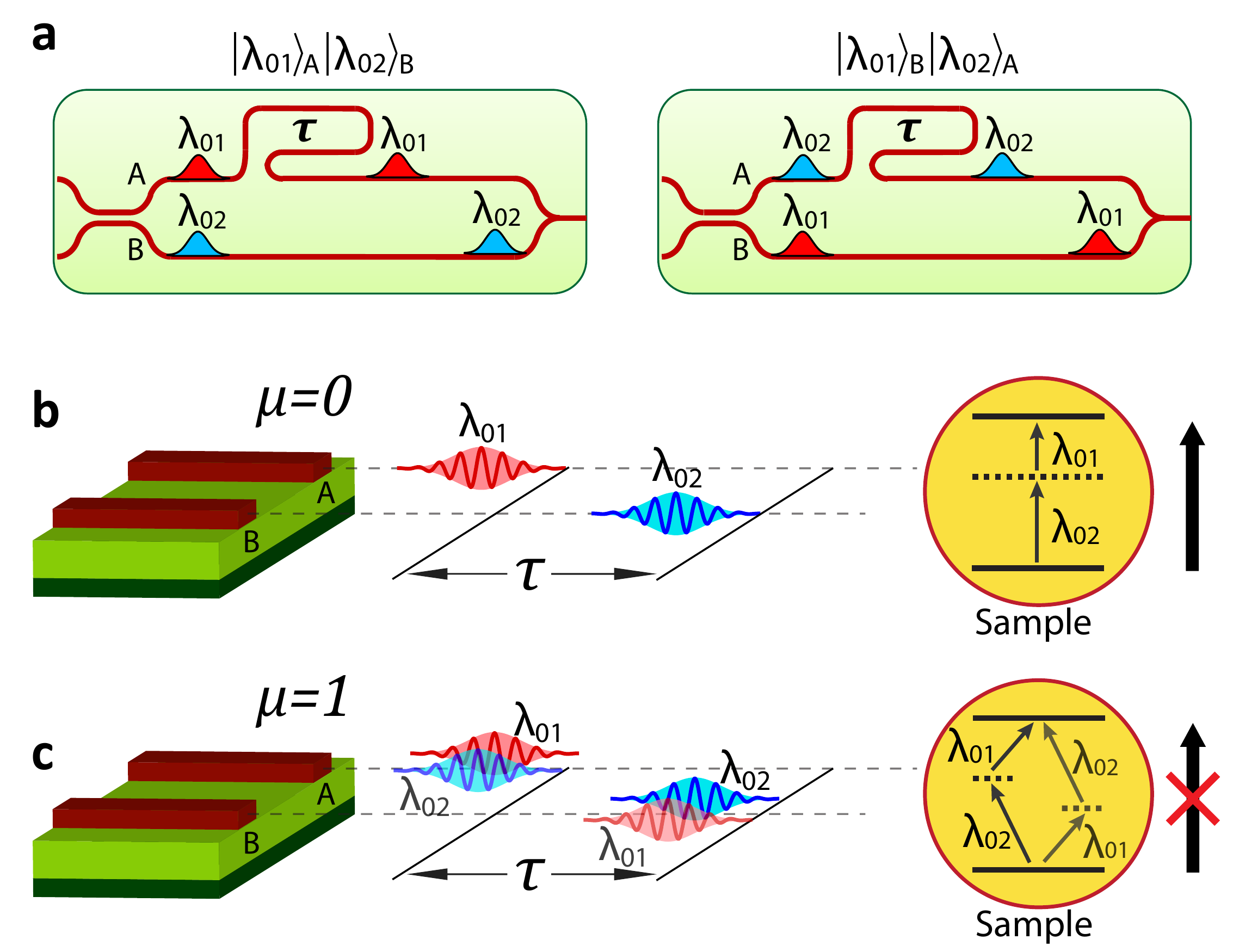}
\caption{\small \textbf{Probing matter with tunable time-ordering.} (\textbf{a}) Photons leaving the coupler from different output ports have two possible pathways: $\vert \lambda_{01}\rangle_{A} \vert \lambda_{02} \rangle_{B}$ or $ \vert \lambda_{01}\rangle_{B} \vert \lambda_{02} \rangle_{A}$. These coincide temporally and hence are mutually coherent. The photon in waveguide $A$ is then temporally delayed by an interval $\tau$ relative to its twin photon in waveguide $B$, so that one photon always arrives at the sample before the other. The wavelength of the delayed photon depends on whether the pathway was  $\vert \lambda_{01}\rangle_{A} \vert \lambda_{02} \rangle_{B}$ or $ \vert \lambda_{01}\rangle_{B} \vert \lambda_{02} \rangle_{A}$. (\textbf{b})~For $\mu = 0$, only the $\vert \lambda_{01}\rangle_{A} \vert \lambda_{02} \rangle_{B}$ pathway is allowed, such that the photon of wavelength $\lambda_{02}$ is always absorbed first. (\textbf{c})~For $\mu = 1$, the superposition permits two absorption pathways: $\lambda_{02}$ followed by $\lambda_{01}$, and $\lambda_{01}$ followed by $\lambda_{02}$. In certain systems~\cite{Muthukrishnan_PRL_2004} where it is not possible to distinguish which of these pathways led to the final state of the sample, the pathways destructively interfere to suppress the two-photon absorption probability. Note that at $\mu = 1$ the pathways $\vert \lambda_{01}\rangle_{A} \vert \lambda_{02} \rangle_{A}$ and $ \vert \lambda_{01}\rangle_{B} \vert \lambda_{02} \rangle_{B}$ are also present due to non-deterministic separation (the coupler behaves as a beamsplitter rather than a WD), which yield photons with no relative delay. These are not time-ordered but do support both absorption pathways and therefore compliment the path-interference effects.}
\label{Fig:TimeOrdering}
\end{figure}

\begin{figure*}[t!]
\centering
\includegraphics[width=2\columnwidth]{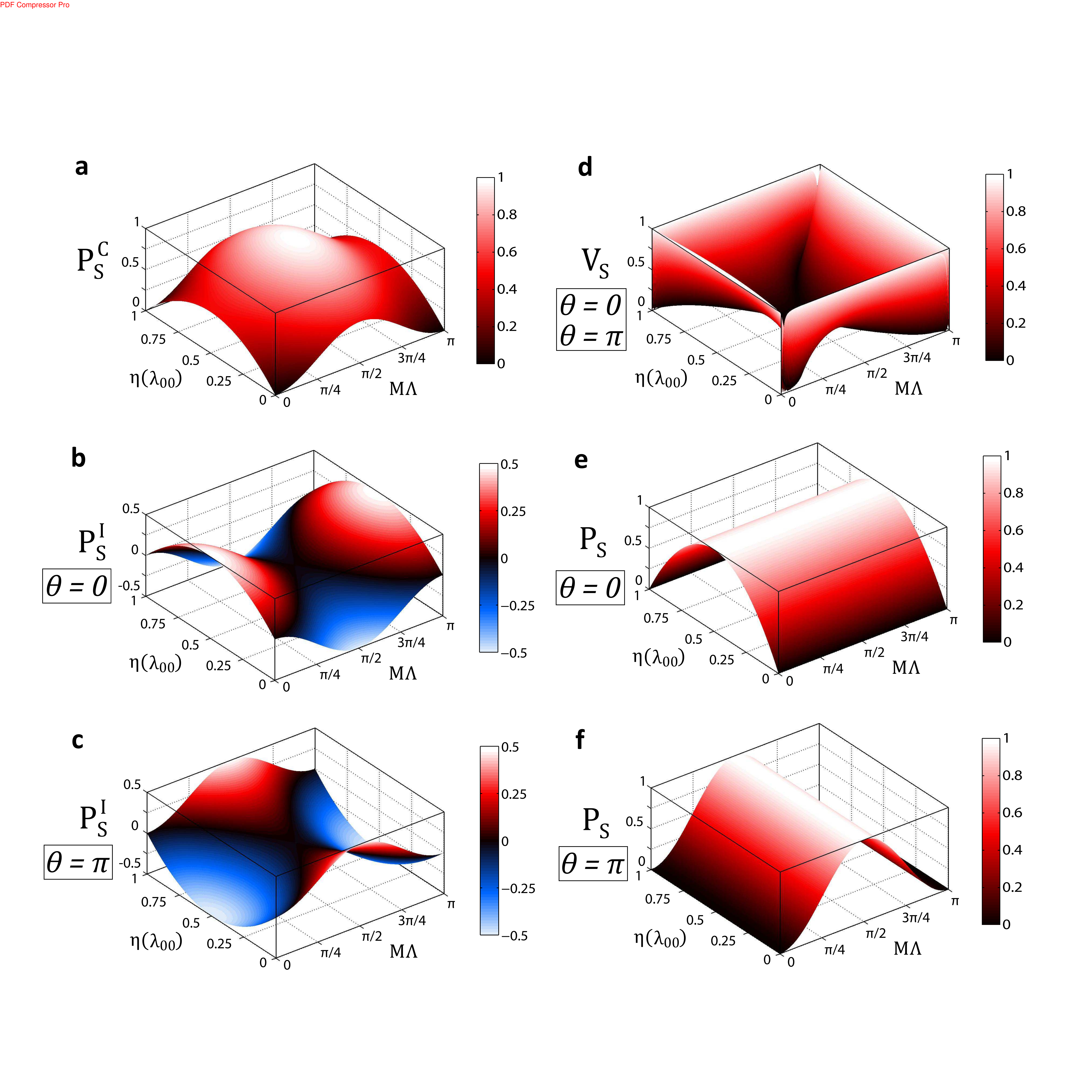}
\caption{\small \textbf{Dependence of two-photon path correlations on coupler response.} Calculations depict (\textbf{a}) the `classical' separation probability, (\textbf{b-c}) the contribution of quantum interference, (\textbf{d}) the resultant interference visibility and (\textbf{e-f}) total separation probability. Toggling the phase shift from $\theta = 0$ to $\theta = \pi$ leads to a sign change for $P_{\textrm{S}}^{\textrm{I}}$ but leaves its magnitude $\vert P_{\textrm{S}}^{\textrm{I}} \vert$ unaltered. This sign change, in turn, toggles the line of maximal $P_{\textrm{S}}$ between $\eta(\lambda_{00}) = 0.5$ and $\mathrm{M}\Lambda = \pi/2$ respectively.}
\label{Fig:PathCorrelations}
\end{figure*}

\subsection{Perfect Anti-Coalesence with Tunable  Visibility}

\par Control over two-photon path correlations is another important ability for quantum photonics. In this section, we start by exploring how such correlations can be impacted by dispersion. We then describe how this enables conditions with no bulk-optics equivalent; namely perfect photon anti-coalescence that remains independent of the visibility of interference effects, even as this visibility is tuned via $\eta(\lambda_{00})$ or $\mathrm{M}\Lambda$. Later in Section~2c, we will highlight possible applications for the dispersion-unlocked correlation behaviour which underscore the breadth of capabilities a single dispersive coupler can provide.  

\par Path correlations are commonly engineered using quantum interference \cite{Hong_PRL_1987, Pittman_PRL_1996, Strekalov_PRA_1998, Chen_PRA_2007,Shadbolt_NaturePhotonics_2011, Silverstone_NaturePhotonics_2014}. In the famous Hong-Ou-Mandel effect \cite{Hong_PRL_1987}, two photons enter a 50:50 beamsplitter from different input paths (anti-bunched), and coalesce to exit as a bunched state where they are most likely to be found in the same output path. Ideally the anti-bunched (i.e. separated) outcome probability becomes $P_{\textrm{S}}=0$ under conditions of maximal interference, compared to the `classical' value of $P_{\textrm{S}}^{\textrm{C}}=0.5$ if interference were completely absent. The reverse process, called anti-coalescence wherein $P_{\textrm{S}} \rightarrow 1$, is useful for providing interference-facilitated pair separation (IFPS) to separate photons generated by integrated sources \cite{Chen_PRA_2007, Silverstone_NaturePhotonics_2014, Jin_PRL_2014}. Note that the subscript $S$ is used to delineate these from probabilities corresponding to bunched (i.e. non-separated) outcomes; this is detailed further in the Methods Section. The two-photon interference can be quantified by the interference visibility $V_{\textrm{S}} = \vert P_{\textrm{S}}^{\textrm{I}} \vert/P_{\textrm{S}}^{\textrm{C}}$, where $P_{\textrm{S}}^{\textrm{I}}  = P_{\textrm{S}} - P_{\textrm{S}}^{\textrm{C}}$ represents the contribution of quantum interference towards the anti-bunched outcome probability.

\par We shall now look specifically at anti-coalescence. While perfect coalescence requires $V_{\textrm{S}}$ to be unity (see Methods), coupler dispersion can lift this restriction for anti-coalescence. As we shall see, for the first time $V_{\textrm{S}}$ can be made to have any arbitrary value between 0 and 1 while the separation probability is kept constant at $P_{\textrm{S}}=1$. Anti-coalescence requires a path-entangled input state of the form
\begin{equation}
\left\vert \Psi \right\rangle = \big[\vert \psi \rangle_{A} \vert 0\rangle_{B} + e^{-i\theta} \vert 0 \rangle_{A}\vert \psi \rangle_{B}  \big] /\sqrt{2},
\label{Eqn:NOON_InputState}
\end{equation}
where $\vert 0 \rangle$ refers to vacuum, $ \left\vert \psi \right\rangle_{j}$ represents a photon pair in path $j$, and $\theta$ is a relative phase shift. Such states can be generated by coherently pumping two sources of photon pairs, as seen in Refs.~\cite{Silverstone_NaturePhotonics_2014} and \cite{Chen_PRA_2007}. This places no restrictions on the tunability of the photon pair sources. The spectral properties of $\vert \psi \rangle_{j}$ are described by the biphoton amplitude (BPA) $\phi^{j}(\omega_{1}, \omega_{2})$. We will assume perfect path indistinguishability such that $\phi^{A}(\omega_{1}, \omega_{2}) = \phi^{B}(\omega_{1}, \omega_{2}) \equiv \phi(\omega_{1}, \omega_{2})$.

\par Figure~\ref{Fig:PathCorrelations} shows how $P_{\textrm{S}}$, $P_{\textrm{S}}^{\textrm{C}}$, $P_{\textrm{S}}^{\textrm{I}}$, and $V_{\textrm{S}}$ change as a function of the coupler parameters, when the relative phase shift is either $\theta = 0$ or $\theta = \pi$. These plots have been generated for a co-polarized input state from Type-I SPDC having $\Delta\lambda = \textrm{0.25~nm}$, $\Delta\lambda_{\textrm{P}} = \textrm{0.1~nm}$ and a degeneracy wavelength of $\lambda_{00} = \textrm{780~nm}$ (see Methods). The dependence of these on photon bandwidth will be discussed in Section~2c. The value of $\vert P_{\textrm{S}}^{\textrm{I}} \vert$ is maximal at coordinates where the coupler responds as a 50:50 beamsplitter, and minimal when it responds as a WD. The `classical' probability $P_{\textrm{S}}^{\textrm{C}}$ follows roughly the opposite trend, obtaining its maximal value of $P_{\textrm{S}}^{\textrm{C}}=1$ for a WD-like response, and decreasing to $P_{\textrm{S}}^{\textrm{C}}=0.5$ for beamsplitter-like responses. Curiously, along the lines $\eta(\lambda_{00}) = 0.5$ and $\mathrm{M}\Lambda = \pi/2$, changes to $P_{\textrm{S}}^{\textrm{C}}$ and $\vert P_{\textrm{S}}^{\textrm{I}} \vert$ are in perfect balance such that their sum always equals unity. This balancing is associated with the condition $\eta(\lambda_{01}) + \eta(\lambda_{02}) = 1$, which leads to $P_{\textrm{S}}= 1$ and hence perfect anti-coalescence (i.e. deterministic separation) along either $\eta(\lambda_{00}) = 0.5$ or $\mathrm{M}\Lambda = \pi/2$, selected through the choice of $\theta$. Along these two lines, the interference visibility $V_{\textrm{S}}$ varies smoothly between $0$ and $1$. By operating at $\mathrm{M}\Lambda = \pi/2$ with $\theta = \pi$, and actively controlling $\eta(\lambda_{00})$ through thermal or electro-optic tuners, any value of $V_{\textrm{S}}$ can be selected while maintaining a perfect separation fidelity. Note that unlike before, this does not alter the spectral entanglement of post-selected output states, due to the presence of path entanglement at the input. 

\par We have just described how coupler dispersion enables the possibility of tuning $V_{\textrm{S}}$ while maintaining $P_{\textrm{S}}=1$. The applications of this capability are not yet known, but its novelty warrants further exploration. It also serves as an example of how integrated components, through their inherent dispersive properties, can access behaviours that bulk bench-top components cannot. Further to this, we now highlight other features of the dispersion-unlocked behaviour that have potential applications for state characterization. 

\subsection{Opportunities for State Characterization}

\subsubsection*{Entanglement-Sensitive Coincidence Detection}
\par For most permutations of coupler and state attributes, $P_{\textrm{S}}$ is accurately described by the behaviour in Figure~\ref{Fig:PathCorrelations}. However, deviations from the values of $P_{\textrm{S}}$ shown can occur when the dimensionless product $\textrm{M}\Delta\lambda$, involving coupler dispersion and photon bandwidth, becomes large. These are described in full at the end of this section. Figure~\ref{Fig:EnanglementSensitivity} indicates that the extent of these deviations depends not only on $\textrm{M}\Delta\lambda$, but also on the spectral entanglement of the input state. This opens up the possibility of discerning the Schmidt Number of the input state from the anti-bunched coincidence count rate at the coupler output, which is proportional to $P_{\textrm{S}}$.

\par The results in Figure~\ref{Fig:EnanglementSensitivity} were calculated for degenerate input states having $\Lambda=\textrm{0~nm}$, $\Delta\lambda = \textrm{10~nm}$, $\lambda_{00} = \textrm{780~nm}$, and $\theta = 0$. The product $\textrm{M}\Delta\lambda$ was swept by varying $\mathrm{M}$, with $\eta(\lambda_{00})=0.5$ kept constant. Input state entanglement was controlled through the Type-I SPDC pump bandwidth $\Delta\lambda_{\textrm{P}}$ (see Methods). In the limit of $\mathrm{M}\Delta\lambda \rightarrow 0$, the above calculation parameters give $P_{\textrm{S}}=1$, in agreement with Fig.~\ref{Fig:PathCorrelations}(e). Larger values of $\textrm{M}\Delta\lambda$ lead to decreases in $P_{\textrm{S}}$. However, increasing the SN of the input state has the effect of asymptotically restoring $P_{\textrm{S}}$ to unity.

\begin{figure}[t!]
\centering
\includegraphics[width=\linewidth]{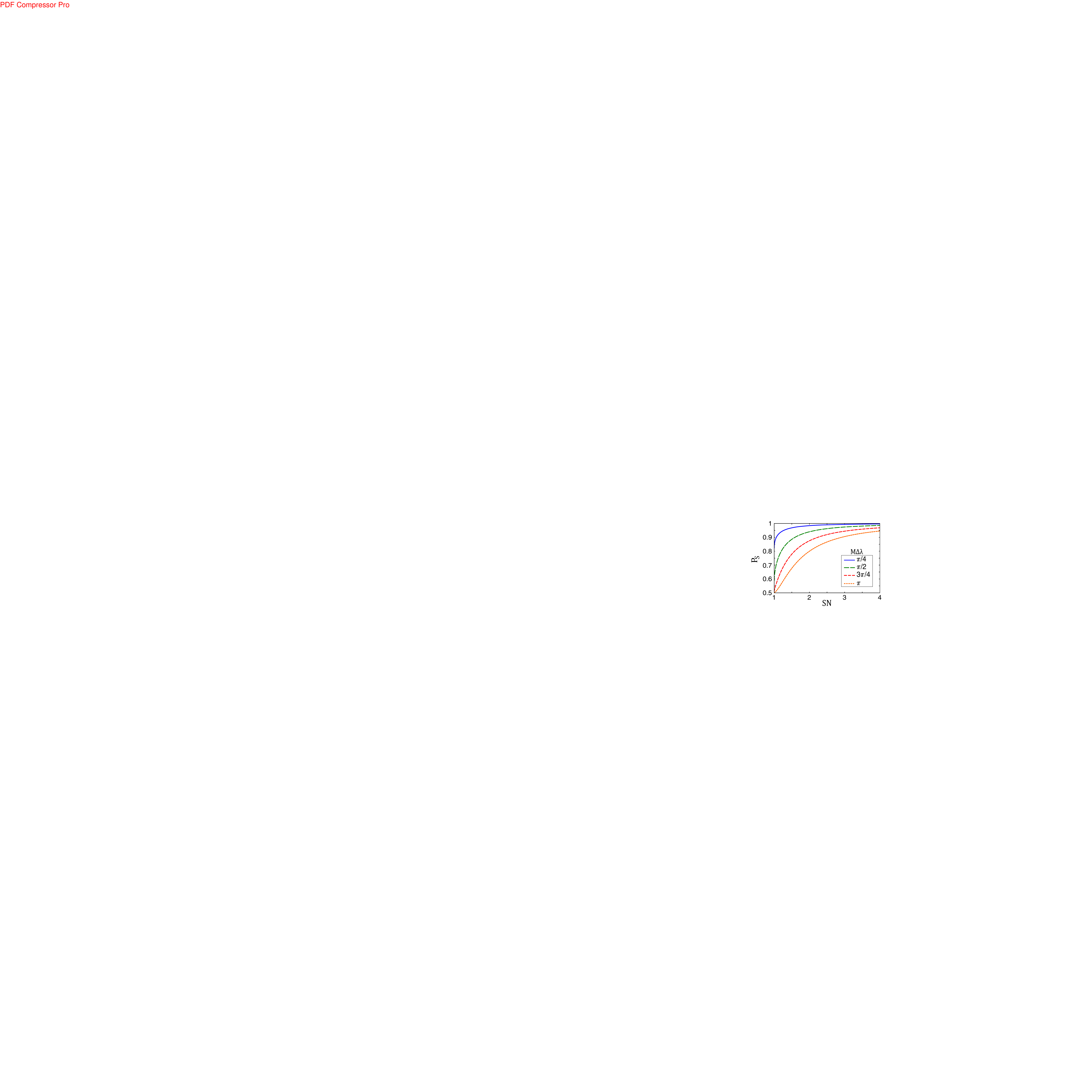}
\caption{\small \textbf{Dependence of $P_{\textrm{S}}$ on entanglement.} The calculated two-photon separation probability is shown as a function of the input state Schmidt Number for $\mathrm{M}\Lambda = 0$, $\eta(\lambda_{00})=0.5$, $\theta = 0$, and $\Delta\lambda = \textrm{10~nm}$, at several values of $\mathrm{M}\Delta\lambda$. For $\textrm{SN} >4$ (not shown), each curve asymptotically approaches unity.}
\label{Fig:EnanglementSensitivity}
\end{figure}

\par This behaviour can be understood by examining Equations~(\ref{Eqn:QuantumCouplerTransformation})-(\ref{Eqn:P_S}) in the Methods section. The probability $P_{\textrm{S}}$ is determined from a sum over all possible combinations of frequencies $\omega_{1} = 2\pi c/\lambda_{1}$ and $\omega_{2}= 2\pi c/\lambda_{2}$ weighted by the BPA. When the state is spectrally uncorrelated (i.e. $\textrm{SN} = 1$), the combinations of $\eta(\lambda_{1})$ and $\eta(\lambda_{2})$ contributing to this sum are not necessarily equidistant from $\eta(\lambda_{00}) = 0.5$ and hence can deviate from the $\eta(\lambda_{1}) + \eta(\lambda_{2}) = 1$ condition required for perfect anti-coalescence. However, when the photons are spectrally anti-correlated due to entanglement, the BPA restricts all contributing $\lambda_{1}$,$\lambda_{2}$ combinations to be approximately equidistant from $\lambda_{\textrm{00}}$, which acts to restore the splitting ratio antisymmetry. Larger products of $\textrm{M}\Delta\lambda$ allow $P_{\textrm{S}}$ to be more severely degraded, because a greater proportion of the non-vanishing $\lambda_{1}$,$\lambda_{2}$ combinations are able to violate the anti-symmetry. Only in the limit of $\Delta\lambda \rightarrow 0$, where the state is entirely described by the central wavelengths $\lambda_{01}$ and $\lambda_{02}$, is the splitting ratio anti-symmetry condition strictly enforced. 

\begin{figure}[t!]
\centering
\includegraphics[width=\linewidth]{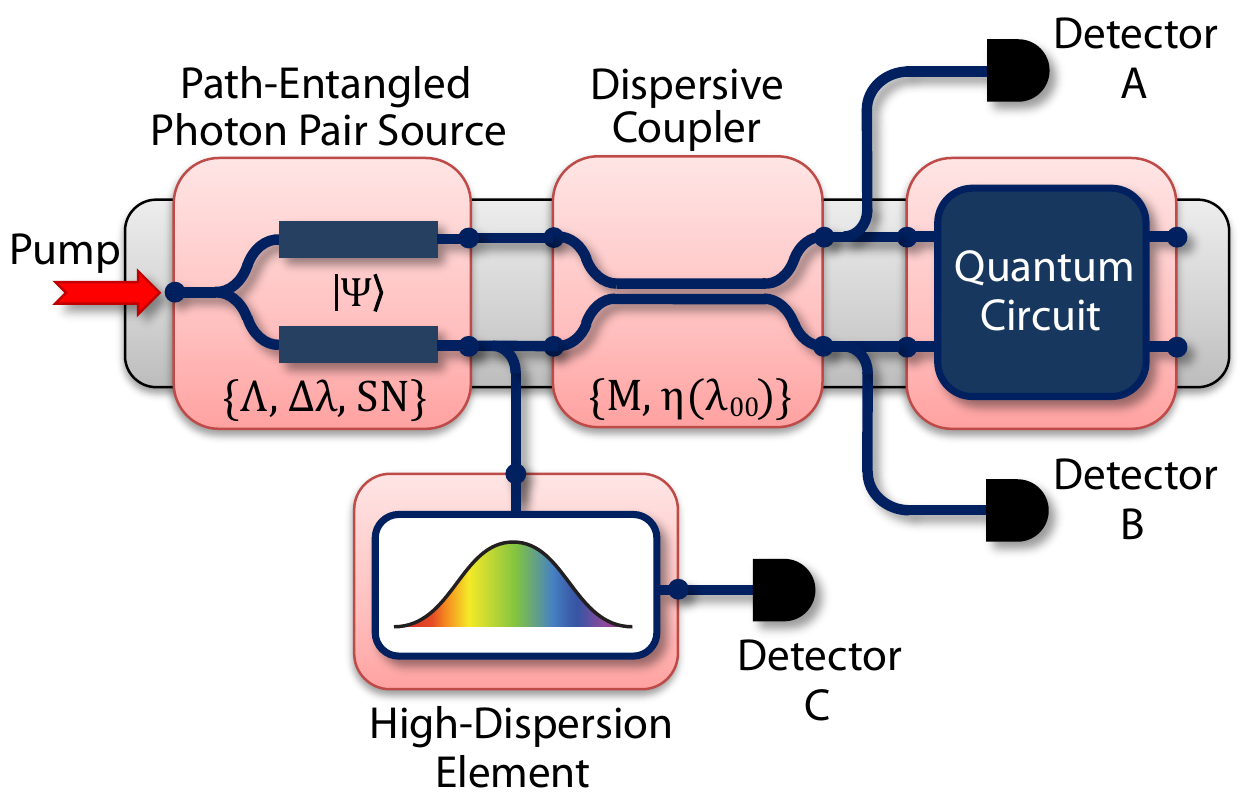}
\caption{\small\textbf{All-integrated $\textrm{SN}$ measurement.}  To apply the technique, the photon pairs must be in the generic path-entangled state $\vert \Psi \rangle$ of Eqn.~(\ref{Eqn:NOON_InputState}). The relative phase is ideally $\theta=0$; for other values of $\theta$, $P_{\textrm{S}}$ is less sensitive to $\textrm{SN}$. To measure $\textrm{SN}$, the state is sampled at three locations (shown as Y-junctions for simplicity). Detectors A and B sample the two-photon statistics at the coupler output to obtain $P_{\textrm{S}}$. Detector C obtains spectrographs, and hence $\Lambda$ and $\Delta\lambda$, by sampling $\vert \Psi \rangle$ via a high-dispersion element such as a fiber or a waveguide grating operated near its band edge. It is sufficient to measure these spectrographs from only one of the source output paths, since the photon pair properties are assumed to be  path-indistinguishable (i.e. $\vert\psi\rangle_{A} = \vert\psi\rangle_{B}$). The data obtained for $\Lambda$ and $\Delta\lambda$ (together with the dispersive coupler attributes) can then be used to map the measured $P_{\textrm{S}}$ to a corresponding value of $\textrm{SN}$ (see Fig.~\ref{Fig:EnanglementSensitivity}).}
\label{Fig:SN_Meas_Example}
\end{figure}

\par The bandwidth and entanglement sensitivity of $P_{\textrm{S}}$ grants dispersive couplers additional capabilities for state characterization. For example, dispersive couplers could empower a simple, fast, all-integrated technique for measuring the Schmidt number of an ensemble of states without needing to perform full state tomography to reconstruct the BPA. Figure~\ref{Fig:SN_Meas_Example} describes how this can be implemented. In this case we show the photons being characterized immediately after leaving the source, in the context of source calibration. However, they could also be measured after interacting with a bath or system. This could be helpful, for example, in metrological applications where the Schmidt number is monitored as an indication of state purity and hence the interaction under investigation. To obtain $\textrm{SN}$, first the marginal photon spectra are measured with a waveguide-assisted spectrograph method \cite{Avenhaus_OL_2009} that uses chromatic group velocity dispersion (GVD) to map spectral components to time-of-arrival at a single-photon detector. Next, provided $\textrm{M}$ is known, the values of $\Delta\lambda$ and $\Lambda$ measured in the first step are used to discern $\textrm{SN}$ from standard two-photon coincidence measurements at the coupler outputs. The sensitivity of the technique diminishes as the photons are made narrowband or increasingly entangled, but can be enhanced by designing the coupler to have $\mathrm{M}$ as large as possible. 

\par Obtaining $\textrm{SN}$ by previous methods would require a measurement of the full BPA, which hinges on the spectral resolution of the measurement system. Measuring the BPA entirely on-chip is possible using spectrographs \cite{Avenhaus_OL_2009}, but its resolution can be severely limited by detector timing jitter. In comparison, precise values of $\Delta\lambda$ and $\Lambda$ for the coupler-assisted technique are more easily obtained, in part due to the straightforward use of interpolation to increase confidence in these values, but also because uncertainties from the limited spectral resolution enter only in one axis, as opposed to two. Hence, the trade-offs between the number of measurements, the total measurement time, and precision in $\textrm{SN}$ scale more favourably for the coupler-assisted technique. A direct, rapid and precise measurement of $\textrm{SN}$ would be particularly useful for the real-time monitoring of sources where $\textrm{SN}$ is tunable \cite{Kumar_NatureComms_2014} and is being used as a control parameter \cite{Schlawin_PRA_2014}. Additionally it would be advantageous for monitoring a stream of states whose properties reveal real-time information about a dynamic system or environment.

\par The converse functionality -- estimating the photon bandwidth for a known Schmidt Number -- could also be useful, in the context of indistinguishable pure photons having tunable attributes \cite{Jin_OE_2013,Shi_OL_2008}. So long as SN remains reasonably close to unity, $\Delta\lambda$ could be measured entirely on-chip using only the coupler and coincidence detectors, without need for tunable bandpass filters, GVD fibers, or spectrometer capabilities. Presently, highly bandwidth-tunable pure photons can be generated in a free-space setup \cite{Shi_OL_2008}, but recent trends towards integration suggest that this capability may eventually be available in a monolithic platform, where on-chip characterization would be helpful for source calibration and monitoring drift. 

\par For completeness, we now return to Figure~\ref{Fig:PathCorrelations} and briefly describe how it changes with bandwidth. When the product $\mathrm{M}\Delta\lambda$ increases but spectral entanglement remains low (i.e. $\textrm{SN}\approx 1$), Figs.~\ref{Fig:PathCorrelations}(a)-(f) all begin to flatten. In Fig.~\ref{Fig:PathCorrelations}(a), the classical contribution $P_{\textrm{S}}^{\textrm{C}}$ at all coordinates approaches a value of 0.5; the interference contributions $P_{\textrm{S}}^{\textrm{I}}$ and visibility $V_{\textrm{S}}$ in Figs.~\ref{Fig:PathCorrelations}(b)-(d) all approach zero; correspondingly the total separation probability $P_{\textrm{S}}$ approaches 0.5 in Figs.~\ref{Fig:PathCorrelations}(e)-(f). In comparison, when the photons are highly frequency-entangled, increases to $\mathrm{M}\Delta\lambda$ do not flatten the surfaces uniformly in this way. Instead, for Figs.~\ref{Fig:PathCorrelations}(a)-(d) it causes the surfaces to 'smear' along the $\mathrm{M}\Lambda$ axis, with the effect of averaging the values along this axis. Figs.~\ref{Fig:PathCorrelations}(e)-(f) are exceptions: for large values of $\textrm{SN}$, $P_{\textrm{S}}$ at $\theta = 0$ remains relatively unchanged from its values at small bandwidths; however,  $P_{\textrm{S}}$ at $\theta = \pi$ instead flattens to approach values of 0.5. These differ because the smearing of Figs.~\ref{Fig:PathCorrelations}(a)-(d) along the $\mathrm{M}\Lambda$ axis alters the symmetry in how the $P_{\textrm{S}}^{\textrm{C}}$ and $P_{\textrm{S}}^{\textrm{I}}$ contributions sum between the two cases.

\subsubsection*{The Versatility of Dispersive Couplers}

\begin{figure}[b!]
\centering
\includegraphics[width=\linewidth]{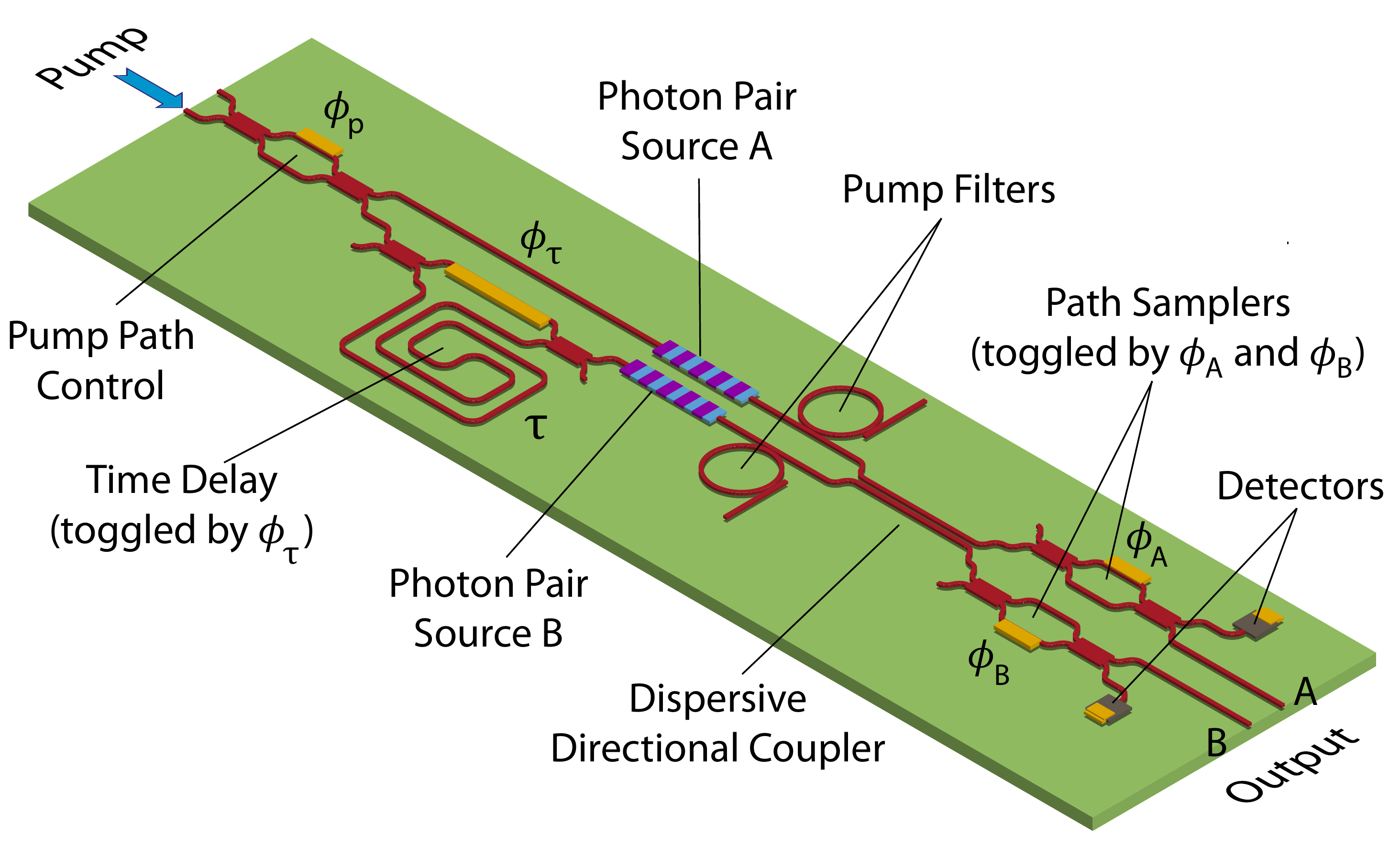}
\caption{\small \textbf{State characterization with a multipurpose dispersive coupler.} A path superposition of the form $\left\vert \Psi \right\rangle$ (Eqn.~\ref{Eqn:NOON_InputState}) is created through coherent pumping of two waveguide sources of photon pairs (e.g. generated via parametric downconversion \cite{Jin_PRL_2014}). A tunable Mach-Zehnder interferometer (MZI) allows the relative time delay to be set to either zero ($\phi_{\tau}=0$) or $\tau$ ($\phi_{\tau}=\pi$). Pump power can be adjusted between paths via $\phi_{p}$ to compensate for asymmetric losses when the delay of $\tau$ is implemented. Unconverted pump photons are removed using ring filters.  MZIs at the output can be toggled ($\phi_{A(B)}=\pi$) to sample the two-photon correlations with single-photon detectors. The rate of detection coincidences for zero time delay and a delay of $\tau$ can be used to determine $V_{\textrm{S}}$, which in turn reveals $\textrm{M}\Lambda$. The dispersive directional coupler must have $\eta(\lambda_{00})=1/2$ for this measurement. Note that adding electro-optic or thermal tuners to the dispersive coupler can enable arbitrary control over $V_{\textrm{S}}$ by tuning $\eta(\lambda_{00})$. Spectral-entanglement tuning is also possible when $\phi_{\textrm{p}}$ is set to deliver pump power to only one of the two photon pair sources.}
\label{Fig:App2_Circuit}
\end{figure}

\par Since couplers are already an essential on-chip device, the state characterization capabilities granted to them by dispersion can be exploited with minimal increase to the circuit complexity or footprint. This allows dispersive couplers to provide an extremely versatile set of functionalities in a compact form factor, which the following example highlights. Consider the reconfigurable circuit in Figure~\ref{Fig:App2_Circuit}. The dispersive coupler in this circuit can serve several purposes. It can provide IFPS to deterministically separate the photons at the coupler output. With the addition of electro-optic or thermal tuning, it can also be utilized for other previously-described state engineering functionalities, such as tunable spectral entanglement. On top of this, the circuit could easily be modified for coupler-based $\textrm{SN}$ measurements by tapping photon source B with a high-dispersion element and additional detector as per Fig.~\ref{Fig:SN_Meas_Example}. Accomplishing all of these tasks through a single dispersive coupler may help make most efficient use of precious on-chip real estate. 

\par Even without adding a tap to source B for a spectrograph measurement, the circuit in Figure~\ref{Fig:App2_Circuit} can already access some information about the state. The relationship between $V_{\textrm{S}}$ and $\Lambda$ described in Section~2b provides a route for measuring the non-degeneracy $\Lambda$ of an ensemble of states entirely on-chip. This requires the toggling of a time delay $\tau$ between the dispersive coupler input paths. The interference visibility is obtained from $V_{\textrm{S}} = \left\vert R_{0}/R_{\tau} - 1 \right\vert$, where $R_{0}$ is the coincidence count rate at zero time delay (as measured by on-chip single photon detectors), and $R_{\tau}$ is the coincidence rate at a time delay $\tau$ that is much larger than the two-photon coherence time (see Methods). Provided $\textrm{M}$ is known, this value of $V_{\textrm{S}}$ can be mapped back to the state non-degeneracy $\Lambda$ as per Fig.~\ref{Fig:PathCorrelations}(d). This technique is best applied to narrow-band photons since the sensitivity of $V_{\textrm{S}}$ to $\Lambda$ decreases as $\mathrm{M}\Lambda$ becomes large.

\section{Dispersive Coupler Example} 

\par We now provide a realistic example of a directional coupler with sufficient dispersion to achieve the capabilities described above. Our aim is to affirm that high dispersion is obtainable under feasible conditions, using a numerically simulated device. The design is intentionally simplistic to show this can be accomplished without much deviation from conventional coupler designs. More optimal approaches will then be discussed. 

\par We consider the manipulation of photon pairs degenerate at 1550~nm in the telecom band, having a maximum tunable non-degeneracy of at least $\Lambda = \textrm{50~nm}$. Such states can be generated through waveguide-based SPDC  (e.g. see Ref.~\cite{Horn_SciRep_2013}). We seek a coupler that can reach the operating point $\mathrm{M}\Lambda = \pi/2$ within this tunable range. 

\par The design of the coupler is shown in Figure~\ref{Fig:Coupler_Example}(a) and is straightforward to fabricate. Figure~\ref{Fig:Coupler_Example}(b) shows its coupling strength in the vicinity of 1550~nm, which is linear and described by $\kappa(\lambda) = 2.2055 \times 10^{10} \lambda - 2.0245 \times 10^{4} \textrm{ ~m\textsuperscript{-1}}$. For 50:50 splitting at the degeneracy point, the smallest suitable interaction length is $L=56.3 \textrm{~{\textmu}m}$. Using the definition of $\textrm{M}$ at the end of Section~1, this yields $\mathrm{M}\Lambda = 0.0621 \approx \pi/50$ at the maximum non-degeneracy of $\Lambda = \textrm{50~nm}$, which is below our target.  However, since $\textrm{M}$ scales linearly with $L$, we can multiply the dispersion by choosing a larger value of $L$ that still gives 50:50 splitting at degeneracy. An interaction length of $L = 1521\textrm{~{\textmu}m}$ achieves this and gives $\mathrm{M}\Lambda = 1.07 \times \pi/2$ for our design, meeting our objective. 

\begin{figure}[h]
\centering
\includegraphics[width=\linewidth]{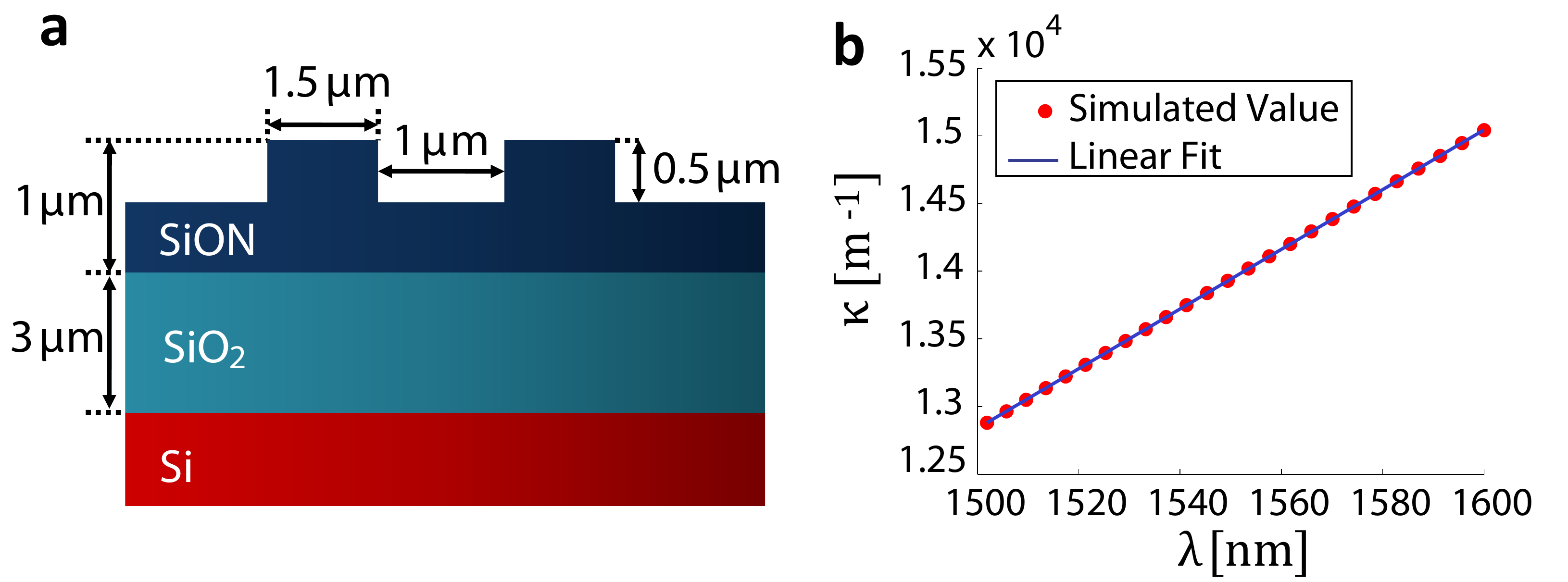}
\caption{\small\textbf{A simple dispersive coupler.} (\textbf{a}) Cross-section of coupler within its coupling length $L$. The design is based on silicon nitride waveguides. (\textbf{b}) Simulated coupling strength $\kappa$ for this design, obtained from commercial FEA software.}
\label{Fig:Coupler_Example}
\end{figure}

\par While the approach above shows that the dispersion can be made arbitrarily large by increasing the device length, this comes at the price of increasing its footprint and insertion losses. Typical losses at 1550~nm for this silicon-nitride waveguide geometry are around 3dB/cm, and hence roughly 10\% of the photons would be lost in the design we described. However, this serves merely as an illustrative example. More sophisticated coupler designs exhibiting appreciable dispersion have been studied in the past. Some examples are grating assisted couplers \cite{Marcuse_JLT_1987,Griffel_JQE_1991} and couplers implemented in asymmetric vertical structures \cite{Peschel_APL_1995}, including Bragg Reflection waveguides \cite{West_OE_2006}. These can provide more compact and efficient ways of achieving the necessary dispersion.

\section{Conclusions}

\par Integrated couplers are already becoming a key building block of photonic quantum circuits. This is partly because they offer greater stability and scalability than bulk-optics beamsplitters and other bench-top components. It is also because the highly precise micron-scale fabrication of such couplers helps eliminate path-length mismatches and other path asymmetries, which is critical for achieving high-fidelity quantum interference \cite{Laing_APL_2010}. However, in addition to these known benefits, our work has revealed an as-of-yet untapped potential for integrated couplers to be utilized in a more versatile way, far beyond their traditional role as a beamsplitter substitute.

\par We found that harnessing the full dispersion properties of an integrated directional coupler unlocks many novel capabilities for the device. These include tunable photon entanglement and time-ordering, as well as bandwidth-sensitive and entanglement-sensitive two-photon effects that can be exploited for state characterization. Some of these capabilities can be achieved in bulk-optics, but not with the convenience nor stability that this integrated approach provides. Yet others have no bulk-optics counterpart, such as the ability to fully tune the two-photon interference visibility (i.e. the sensitivity to time delays at the coupler input) while maintaining a constant flux of separated (i.e. anti-coalesced) photon pairs. Particularly remarkable is that all of these functionalities can be provided by a single integrated coupler, making it a versatile yet compact tool for both state engineering and on-chip state characterization. This is made possible by the capacity of dispersive couplers to smoothly transition between the extremes of beamsplitter and wavelength-demultiplexer behaviour, in a manner without parallel in bulk optics. 

\par Ultimately, we have shown that conventional integrated optics devices can have much more to offer quantum optics if re-evaluated in unconventional ways. Our analysis can be extended to provide a fresh look at several other coupler types, and the additional capabilities their dispersive characteristics might offer. These include multimode interferometers and rings, as well as atypical coupler geometries with more exotic transfer functions \cite{Takagi_JLT_1992}. Cavity-based couplers may also have interesting uses when examined beyond the identical-photon regime \cite{Agarwal_PRA_1994}. This work also lays foundations for studying the implications of dispersion in coupled waveguide arrays \cite{Bromberg_PRL_2010} and three-dimensional tritters \cite{Spagnolo_NatureComms_2013}. Such systems can be considered not only for two-photon phenomena, but also for engineering higher-order path correlation effects involving multi-pair production or multiple sources.

\section*{Methods}
\small

\par \textbf{Parameterization of generalized coupler response.} Our analysis assumes that the waveguides are single-mode, identical, and rectangular (i.e. non-tapered) as illustrated in Figure~\ref{Fig:CouplerResponse}(a), although more sophisticated design geometries are possible \cite{Takagi_JLT_1992}. Spatial mode overlap \cite{Yariv_JQE_1973, Taylor_ProcIEEE_1974} between the waveguides leads to $\eta(\lambda) = \cos^{2}\left(\kappa(\lambda)L\right)$ in terms of a coupling strength $\kappa(\lambda)$ over an interaction length $L$. As defined, $\eta(\lambda)$ represents the probability that a photon exits from the same waveguide it enters from (i.e.  $\eta(\lambda)=1$ means no power is transferred). It is useful to parameterize the coupler's response to the two-photon state in terms of generic dimensionless variables that can be mapped to any combination of coupler and state properties. The dimensionless product $\textrm{M}\Lambda$, where $\Lambda = \vert \lambda_{02} - \lambda_{01} \vert$ is the photon pair non-degeneracy and $\textrm{M} = \mathrm{d}\kappa(\lambda)L/\mathrm{d}\lambda$ is the first-order coupler dispersion, gives the absolute difference in $\kappa(\lambda)L$ between the photon central wavelengths. For discussing spectral dependencies, the product $\textrm{M}\Delta\lambda$ similarly gives the absolute difference in $\kappa(\lambda)L$ across the FWHM of the marginal spectra.

\par A convenient parameter space for navigating the coupler response can be created from $\eta(\lambda_{00})$ and $\mathrm{M}\Lambda$ if the reference wavelength $\lambda_{00}$ is taken to be the average of the photon central wavelengths $\lambda_{01}$ and $\lambda_{02}$. For photon pairs with a tunable non-degeneracy, such as those generated through spontaneous nonlinear interactions \cite{Rubin_PRA_1994, Yang_PRA_2008, Chen_PRA_2005}, $\lambda_{00}$ can be set as the photon pair degeneracy wavelength, since $\lambda_{01}$ and $\lambda_{02}$ tend to remain approximately equidistant from the degeneracy point for $\Lambda$ of up to hundreds of nanometers (e.g. see the pair generation tuning curves of Fig. 2 in Ref.~\cite{Horn_SciRep_2013}). All possible coupler responses to the quantum state then occur within the bounds $\eta(\lambda_{00}) \in [0,1]$ and $\mathrm{M}\Lambda \in [0, \pi]$.  Behaviours for $\mathrm{M}\Lambda > \pi$ can be mapped back to the interval $\mathrm{M}\Lambda \in [0, \pi]$.
Figure~\ref{Fig:CouplerResponse}(b) shows how $\Delta\eta$ varies within these bounds. There are four coordinates where the coupler responds as a 50:50 beamsplitter with $\eta(\lambda_{01})=\eta(\lambda_{02}) = 0.5$, and one central coordinate where it responds as a WD with $\Delta\eta = 1$. These provide a reference for tracking transitions between beamsplitter and WD behaviour. The special condition $\eta(\lambda_{01}) + \eta(\lambda_{02}) = 1$ occurs along the lines $\eta(\lambda_{00})=0$ and $\mathrm{M}\Lambda = \pi/2$, where the splitting ratios at $\lambda_{01}$ and $\lambda_{02}$ are anti-symmetric about $\eta = 0.5$. We note that if the assumptions of linear $\kappa(\lambda)$ or $\lambda_{00} = \vert \lambda_{01} + \lambda_{02} \vert / 2$ break down, the parameter space shown in Figure~\ref{Fig:CouplerResponse}(b) becomes skewed with respect to the horizontal axis.

\vspace{1.5\baselineskip} \par \noindent \textbf{State Representation.}  A co-polarized pair with both photons beginning in waveguide $j$ can be represented by the pure state
\begin{equation}
\left\vert \psi \right\rangle_{j} = \int \mathrm{d}\omega_{1} \mathrm{d} \omega_{2} \, \phi^{j}(\omega_{1},\omega_{2}) \hat{a}^{j \dagger}(\omega_{1}) \hat{a}^{j \dagger}(\omega_{2}) \left\vert \text{vac} \right\rangle,
\label{Eqn:psi_j}
\end{equation}
where $\hat{a}^{j\dagger}(\omega)$ is the canonical mode creation operator for waveguide $j$. 
The BPA is normalized according to $\int \mathrm{d}\omega_{1} \mathrm{d}\omega_{2} \, \big\vert \phi^{j}(\omega_{1}, \omega_{2}) \big\vert = 1$. Rather than generating the BPA from device-specific mode dispersion parameters \cite{Chen_PRA_2005, Yang_PRA_2008}, it is more convenient to define the BPA directly in terms of the photon bandwidths and central wavelengths of interest. A BPA that mimics the output of a Type I SPDC process can be constructed from
\begin{equation}
\phi(\omega_{1}, \omega_{2}) = \phi_{\textrm{P}}(\omega_{1} + \omega_{2})\left[\phi_{1}(\omega_{1})\phi_{2}(\omega_{2})   + \phi_{2}(\omega_{1})\phi_{1}(\omega_{2})  \right],
\label{Eqn:BPA_Construction}
\end{equation}
where $\phi_{n}(\omega)$ are the marginal photon spectra and $\phi_{\textrm{P}}(\omega_{1} + \omega_{2})$ is the pump spectrum. This construction satisfies the necessary exchange symmetry and has all the key qualitative features of a typical Type I BPA computed from SPDC theory. The marginal spectra were gaussian and defined in terms of wavelength as $\phi_{n}(\lambda) = \exp\big( -2 \ln 2 \left[ \lambda - \lambda_{0n} \right]^{2} / \Delta\lambda^{2} \big)$, with equal FWHM intensity bandwidths of $\Delta\lambda$.  The pump spectrum was also gaussian with a FWHM intensity bandwidth of  $\Delta\lambda_{\textrm{P}}$. Narrowing $\Delta\lambda_{\textrm{P}}$ below $\Delta\lambda$ has the effect of increasing the spectral correlations, and hence Schmidt Number, of the two-photon state.

\vspace{1.5\baselineskip} \par \noindent \textbf{Evolution through a directional coupler.} Consider the evolution of the pure state $\left\vert \Psi \right\rangle$ of Equation~(\ref{Eqn:NOON_InputState}) through a directional coupler of length $L$ and coupling strength $\kappa(\omega)$. It is assumed that the output remains in a pure state. Let $\hat{b}^{j}(\omega)$ represent the mode operators at the coupler output. These are related to the input mode operators by
\begin{equation}
\renewcommand{\arraystretch}{1.25} \left[ \begin{array}{c} \hat{b}^{A \dagger}(\omega) \\ \hat{b}^{B \dagger}(\omega) \end{array} \right] = \renewcommand{\arraystretch}{1.25} \left[ \begin{array}{cc} \cos\left(\kappa(\omega)L\right) & i\sin\left(\kappa(\omega)L\right) \\  i\sin\left(\kappa(\omega)L\right)  & \cos\left(\kappa(\omega)L\right) \end{array} \right] \renewcommand{\arraystretch}{1.25}  \left[ \begin{array}{c} \hat{a}^{A \dagger}(\omega)  \\ \hat{a}^{B \dagger}(\omega) \end{array} \right].
\label{Eqn:QuantumCouplerTransformation}
\end{equation}
Note that the magnitude of the matrix elements in Equation~(\ref{Eqn:QuantumCouplerTransformation}) are related to the power splitting ratio by $\left\vert  \cos\left(\kappa(\omega)L\right) \right\vert = [\eta(\omega)]^{-1/2}$  and $\left\vert  \sin\left(\kappa(\omega)L\right) \right\vert = [1 - \eta(\omega)]^{-1/2}$. Using this transformation, the state BPAs at the output of the coupler can be written as follows:
\begin{equation}
\Phi^{j \rightarrow pq}(\omega_{1}, \omega_{2}) = \phi^{j}(\omega_{1}, \omega_{2}) G^{j \rightarrow p}(\omega_{1}) G^{j \rightarrow q}(\omega_{2}),
\label{Eqn:Phi_JPQ}
\end{equation}
where
\begin{equation}
G^{j \rightarrow q}(\omega) = \begin{cases} \cos\left(\kappa(\omega)L\right), & \mbox{if } j = q \\ \sin\left(\kappa(\omega)L\right), & \mbox{if } j \neq q \end{cases}.
\label{Eqn:GreensFunctions}
\end{equation}
In terms of our notation, $\Phi^{j \rightarrow pq}(\omega_{1}, \omega_{2})$ is the amplitude associated with photons 1 and 2 being coupled from input path $j$ to output paths $p$ and $q$ respectively. While the form of Eqn.~(\ref{Eqn:Phi_JPQ}) is general, the $G^{j \rightarrow q}(\omega)$ will change if a different coupler architecture is used (such as an asymmetric coupler).

\vspace{1.5\baselineskip} \par \noindent \textbf{Two-photon outcome probabilities.} The probability of finding photons 1 and 2 in output paths $p$ and $q$ respectively is calculated from $P_{pq} = \left\langle \Psi \right\vert  \hat{b}^{p\dagger}  \hat{b}^{q\dagger}  \hat{b}^{q}  \hat{b}^{p}\left\vert \Psi \right\rangle$ and found to be 
\begin{equation}
P_{pq} = R^{\textrm{C}}_{pq} + \cos( \pi \delta_{pq}) R^{\text{\textrm{I}}}_{pq}(\theta),
\label{Eqn:P_pq}
\end{equation}
where $\delta_{pq}$ is the Kronecker delta, 
\begin{equation}
R^{\textrm{C}}_{pq}  =  \int  \mathrm{d} \omega_{1} \mathrm{d} \omega_{2} \,  \Big( \big\vert \Phi^{A \rightarrow pq}(\omega_{1}, \omega_{2}) \big\vert^{2}  +  \big\vert \Phi^{B \rightarrow pq}(\omega_{1}, \omega_{2}) \big\vert^{2} \Big),
\label{Eqn:R_pq_C}
\end{equation}
is the `classical' probability contributed by sources $A$ and $B$ in the absence of interference, and
\begin{equation}
R^{\textrm{I}}_{pq}(\theta) =  \int  \mathrm{d} \omega_{1} \mathrm{d} \omega_{2}\, 2\text{Re} \Big\{e^{-i\theta} \Phi^{B \rightarrow pq}(\omega_{1}, \omega_{2}) \Phi^{*A \rightarrow pq}(\omega_{1},\omega_{2})  \Big\},
\label{Eqn:R_pq_I}
\end{equation}
is a non-classical modifier accounting for the effects of path interference. These expressions are given in their most general form so that they can be readily applied to any arbitrary set of coupler and two-photon state attributes. Note that $\sum_{pq}P_{pq} = 1$. The probability $P_{\textrm{S}}$ of obtaining an anti-bunched (separated) outcome is then 
\begin{equation}
P_{\textrm{S}} = P_{A B} + P_{B A} = P_{\textrm{S}}^{\textrm{C}} +  P_{\textrm{S}}^{\textrm{I}} 
\label{Eqn:P_S}
\end{equation}
with `classical' and `interference' components given by $P_{\textrm{S}}^{\textrm{C}} = R_{AB}^{\textrm{C}} + R_{BA}^{\textrm{C}}$ and $P_{\textrm{S}}^{\textrm{I}} = R_{AB}^{\textrm{I}} + R_{BA}^{\textrm{I}}$.

\vspace{1.5\baselineskip} \par \noindent \textbf{Calculation of Spectral Entanglement.} The spectral entanglement of a state is completely described by its BPA. For a given BPA, the Schmidt Number is calculated from $\mathrm{SN} = 1 / \left[ \sum_{n} p_{n}^{2} \right]$, where the $p_{n}$ are the eigenvalues of the matrix \cite{Parker_PRA_2000, Humble_PRA_2008}
\begin{equation}
\rho_{\omega \omega'} = \int \mathrm{d} \omega'' \,  \phi(\omega, \omega'') \phi^{*}(\omega', \omega''),
\label{Eqn:SchmidtDecomposition_ MatrixElements}
\end{equation}
and are normalized according to $\sum_{n} p_{n} = 1$. To quantify the entanglement of anti-bunched states at the coupler output,  we associate the labels 1 and 2 with output paths $A$ and $B$ respectively, and post-select for terms containing $\hat{b}^{A\dagger}(\omega_{1})\hat{b}^{B\dagger}(\omega_{2})\vert \textrm{vac} \rangle$. The associated BPA is proportional to
\begin{equation}
\varXi^{AB}(\omega_{1}, \omega_{2}) = \Phi^{A \rightarrow AB}(\omega_{1}, \omega_{2}) + \Phi^{B \rightarrow AB}(\omega_{1}, \omega_{2}),
\end{equation}
which replaces $\phi(\omega_{1}, \omega_{2})$ in Equation~(\ref{Eqn:SchmidtDecomposition_ MatrixElements}). For the non-path-entangled input state $\vert \psi \rangle_{A}$, we set $\Phi^{B \rightarrow AB}(\omega_{1}, \omega_{2})$ to zero.

\vspace{1.5\baselineskip} \par \noindent \textbf{Obtaining $V_{\textrm{S}}$ for on-chip measurement of $\Lambda$.} We refer to the configuration shown in Figure~\ref{Fig:App2_Circuit}. Let $P_{\textrm{S}}(\Lambda,\tau)$ represent the total anti-bunched outcome probability at non-degeneracy $\Lambda$ and relative time delay $\tau$. Assuming $\eta(\lambda_{00})=0.5$ and $\theta = 0$,  $P_{\textrm{S}}(\Lambda,0)=1$ at all values of $\Lambda$. The coincidence detection rate $R_{0}$ at zero delay therefore corresponds to maximum separation fidelity; thus the probability of pair separation at non-zero delay $\tau$ can be obtained from $P_{\textrm{S}}(\Lambda,\tau)=R_{\tau}/R_{0}$. Provided $\tau$ is large enough that $\left\vert\psi\right\rangle_{A}$ and $\left\vert\psi\right\rangle_{B}$ (the possible photon-pair histories) are no longer coherent, quantum interference will not occur at that delay time; thus $P_{\textrm{S}}^{\textrm{I}}(\Lambda,\tau) = 0$ and $P_{\textrm{S}}(\Lambda,\tau)=P_{\textrm{S}}^{\textrm{C}}(\Lambda,\tau)$. It then follows from the definition of $V_{\textrm{S}}$ that 
\begin{equation}
V_{\textrm{S}} = \frac{\left\vert P_{\textrm{S}}(\Lambda,0) -P_{\textrm{S}}(\Lambda,\tau) \right\vert}{P_{\textrm{S}}(\Lambda,\tau)} = \frac{\left\vert 1 - R_{\tau}/R_{0}\right\vert}{R_{\tau}/R_{0}} = \left\vert \frac{R_{0}}{R_{\tau}} - 1\right\vert.
\end{equation} 
For $\eta(\lambda_{00})=0.5$, the visibility $V_{\textrm{S}}$ maps to a unique value of $\mathrm{M}\Lambda$ provided $\mathrm{M}\Lambda \leq \pi/2$ (due to periodicity of $V_{\textrm{S}}$; see Fig.~\ref{Fig:PathCorrelations}(d)).

\vspace{1.5\baselineskip} \par \noindent \textbf{Other Remarks} In addition to the separated (anti-bunched) probabilities $P_{\textrm{S}}^{\textrm{C}}$, $P_{\textrm{S}}^{\textrm{I}}$, and $P_{\textrm{S}}$, there is naturally a complementary set of bunched probabilities $P_{\textrm{B}}^{\textrm{C}}$, $P_{\textrm{B}}^{\textrm{I}}$, and $P_{\textrm{B}}$, corresponding to outcomes where the photons exit together from the same output port. For anti-coalescence, these are related as follows:  $P_{\textrm{S}} +  P_{\textrm{B}} = 1$; $P_{\textrm{S}}^{\textrm{C}} + P_{\textrm{B}}^{\textrm{C}} = 1$; and $\vert P_{\textrm{S}}^{\textrm{I}} \vert = \vert P_{\textrm{B}}^{\textrm{I}} \vert$. It is likewise possible to define a bunched-outcome interference visibility $V_{\textrm{B}} = \vert P_{\textrm{B}}^{\textrm{I}} \vert /  P_{\textrm{B}}^{\textrm{C}} $, which behaves differently from $V_{\textrm{S}}$. 

\par The behaviour of these visibilities also depends on whether we are implementing coalescence (i.e. with photons beginning in different waveguides) or anti-coalescence (i.e. with photons beginning in the same waveguide). For simplicity, consider the familiar case where the coupler is non-dispersive and hence $\eta$ is a fixed value. For coalescence such as in the HOM effect, $V_{\textrm{S}} = 2\eta(1-\eta)/\left[  \eta^2 + (1-\eta)^2 \right]$, while $V_{\textrm{B}} =1$ and is independent of $\eta$ because the classical and non-classical contributions to $P_{\textrm{B}}$ scale identically. These behaviours are reversed for anti-coalescence. We also note that without dispersion, both visibilities must be equal to unity for perfect coalescence or anti-coalescence to occur. However, with dispersion, this requirement is lifted.

\small
\bibliographystyle{osajnl} 
\raggedright
\bibliography{journalTitlesAbbrv,RPM_MasterBib} 

\end{document}